\documentclass[10pt]{article}
\usepackage{titlesec}
\usepackage{amsmath,esdiff}
\usepackage{amssymb}
\usepackage{authblk}

\usepackage[normalem]{ulem}
\usepackage{bbold}
\usepackage[colorlinks=true,linkcolor=blue,citecolor=red,urlcolor=red]{hyperref}%
\usepackage{geometry}
\usepackage{setspace}
\usepackage{BOONDOX-cal}
\usepackage{cite}
\usepackage{flushend}
\usepackage{mathrsfs}
\usepackage{hyperref}
\usepackage{graphicx}
\usepackage{cancel}
\usepackage{array,esint}
\usepackage[most]{tcolorbox}
\graphicspath{}
\titleformat{\section}{\fontsize{12}{12}\bfseries}{\thesection}{1em}{}
\twocolumn
\begin{document}
\twocolumn[\begin{@twocolumnfalse}
\title{\textbf{Interference and reflection from the event horizon of a quantum corrected black hole}}
\author{\textbf{Sunandan Gangopadhyay${}^{a*}$, Soham Sen${}^{a\dagger}$ and Rituparna Mandal${}^{a\ddagger}$}}
\affil{{${}^a$ Department of Astrophysics and High Energy Physics}\\
{S.N. Bose National Centre for Basic Sciences}\\
{JD Block, Sector III, Salt Lake, Kolkata 700 106, India}}
\date{}
\maketitle
\begin{abstract}
\noindent In this work, we calculate the Hawking temperature for a quantum corrected black hole geometry using the \textit{reflection from the horizon} method. We observe that quantum gravity corrections indeed show up in the Hawking temperature formula of the quantum corrected black hole. It is important to notice that the quantum gravity corrections arise in the Hawking temperature formula only due to the underlying quantum gravity corrections to the lapse function of the black hole metric rather than the semi classical methods used in the analysis. We also substantiate our result by computing the Hawking temperature using the tunneling approach. 
\end{abstract}
\end{@twocolumnfalse}]
\section{Introduction}
\footnote{{}\\
{$*$sunandan.gangopadhyay@gmail.com}\\
{$\dagger$sensohomhary@gmail.com, soham.sen@bose.res.in}\\
{$\ddagger$drimit.ritu@gmail.com}}
General theory of relativity developed by Albert Einstein is considered as the most accurate theory describing the large scale structure of the universe\cite{Einstein15,Einstein16}. The Einstein's field equations admit solutions with singularities which are called black holes. In classical consideration any object that falls into the event horizon of the black hole can never escape from it. Therefore classically black holes act as perfect absorbers. This classical picture has known limitations. From the viewpoint of thermodynamics, a black hole should have entropy and a temperature. It was shown in \cite{Christodoulou} that if one tries to extract energy from a Kerr black hole, there is a quantity which never decreases and it was later found to be proportional to the area of the black hole. Future investigations revealed that the area of the black hole is its physical entropy\cite{Bekenstein,Bekenstein2,Bekenstein3}. Some ground breaking works also revealed that the black holes have their own laws of thermodynamics\cite{Hawking0}. It was first shown by Stephen Hawking that if one considers quantum fluctuations in the curved background, the black hole emits radiation\cite{Hawking,Hawking2,Hawking3}. Later it was implemented by Stephen Hawking that the black hole radiation is identical to that of a black body radiation with temperature $T=\frac{\kappa}{2\pi}$, where $\kappa$ is the surface gravity of the black hole. There have been several attempts to give an alternative derivation of the Hawking radiation\cite{HartleHawking,GibbonsHawking,ChristensenFulling}. Some of the very famous approaches to understand the origin of the Hawking temperature revolves around the \textit{reflection from the horizon} method\cite{Kuchiev} and the well known tunneling approach\cite{PWZ,Sripadma}. The radial null geodesic approach has been used later to compute the Hawking temperature for different black hole geometries\cite{Jiang,ErratumJiang,LiuZhu,XuChen,YuZhao}. The tunneling of Dirac particle through the event horizon has also been considered in \cite{KernerMann,DChen,Cricienzo,RLi,KernerMann2,DYChen}.

\noindent It was shown in \cite{Kuchiev} that if a particle is falling into the event horizon of a black hole, there is  a finite probability that the particle will be reflected back  from the event horizon of the black hole. If one considers particle trajectories around the black hole, classically there are ingoing trajectories and the outgoing trajectories. The outgoing trajectories describe the trajectories that leads a particle away from the centre of the black hole and the ingoing trajectories on the other hand leads particles towards the singularity of the black hole. If a particle enters the event horizon of a black hole, there are no ways to escape classically. But if one consider quantum mechanical analysis in the vicinity of the event horizon, it can be observed that the solution of the incoming particle consists of both the ingoing and outgoing trajectories to allow superposition between the two trajectories. This outgoing part of the solution implies a finite probability of a particle to be reflected back from the event horizon of the black hole making the absorption cross-section finite which is proportional to the event horizon area in the infrared region\cite{ZhKaplan,Zel'dovich}. Now the Hawking radiation process includes radiations being emitted from the event horizon of the black hole. It was analyzed in \cite{Kuchiev} that if one calculates the temperature of the black hole for simple black hole geometries, it coincides with the Hawking temperature of the black hole implying that this \textit{reflection from the horizon} can work as an alternative origin story for the Hawking radiation of a general black hole geometry.   

\noindent The tunneling approach serves as an alternate description of the source of the Hawking radiation which describes the source of this radiation by tunneling of scalar waves across the event horizon. The core idea of this approach lies in the fact that there is an abrupt change in sign of the energy of the particle when it crosses the event horizon of a black hole. If a pair is created outside or inside of the event horizon of the black hole, after one member of the pair has tunneled to the opposite site, it can materialize with zero total energy. The mass of the black hole goes down as it radiates to maintain conservation of energy. These black holes are therefore in highly excited states from a quantum gravity point of view. The more popular approach to treat Hawking radiation is by considering the black hole immersed in a thermal bath in which equilibrium is possible. Hence, by the principle of detailed balance there must be emission from the black hole itself.

\noindent In this work, our main motivation is to investigate the Hawking temperature of quantum gravity corrected black hole geometries  via both the \textit{reflection from the horizon} method and the tunneling approach. As we are working with quantum gravitational black holes, one expects to find quantum gravitational corrections to the classical form of the Hawking temperature. Here we have considered the quantum corrected black hole geometry obtained from the flow of the Newton's gravitational constant and the \textit{Garfinkle-Horowitz-Strominger} (GHS) black hole.

\noindent In section 2 we have obtained the scalar field solution of the covariant Klein-Gordon equation using the Hamilton-Jacobi approach. In sections 3 and 4, we have obtained the Hawking temperatures for the quantum corrected black hole and the GHS black hole using the the method of \textit{reflection from the horizon}.
\section{Hamilton-Jacobi method for obtaining the scalar field solution}
In the $s$-wave approximation, the spacetime structure of a spherically symmetric black hole is effectively $1+1$ - dimensional. The generic metric structure for such a static spherically symmetric black hole reads
\begin{equation}\label{1.1}
ds^2=-f(r)dt^2+g(r)^{-1}dr^2~.
\end{equation}
We shall now consider a scalar field in the above background. It satisfies the Klein-Gordon quantum field equation for a scalar field with rest mass $m$ in two spacetime dimensions 
\begin{equation}\label{1.2}
-\frac{\hbar^2}{\sqrt{-g}}\partial_{\mu}\left(\sqrt{-g}g^{\mu\nu}\partial_\nu\right)\Psi=m^2\Psi
\end{equation}    
where $\sqrt{-g}=\sqrt{-\det{(g_{\mu\nu})}}=\sqrt{\frac{f(r)}{g(r)}}$ and $\Psi$ is the scalar field. We will be considering massless scalar fields in our analysis, and using the metric structure in eq.(\ref{1.1}), we can express eq.(\ref{1.2}) as follows
\begin{equation}\label{1.3}
-\frac{\partial_t^2\Psi}{f(r)}+\left(\frac{f'(r)}{2f(r)}+\frac{g'(r)}{2g(r)}\right)g(r)\partial_r\Psi+g(r)\partial_r^2\Psi=0~.
\end{equation}
In order to obtain the solution of eq.(\ref{1.3}), we take an ansatz of the form 
\begin{equation}\label{1.5}
\Psi(t,r)=\exp\left(-\frac{i}{\hbar}I(t,r)\right)~.
\end{equation}
Substituting the ansatz in eq.(\ref{1.5}) for $\Psi(t,r)$ in eq.(\ref{1.3}), we obtain the following equation involving $I(t,r)$
\begin{equation}\label{1.6}
\begin{split}
&\frac{i}{f(r)}\left(\frac{\partial I}{\partial t}\right)^2-\frac{\hbar}{f(r)}\frac{\partial^2 I}{\partial t^2}-i g(r)\left(\frac{\partial I}{\partial r}\right)^2\\&
+\hbar g(r)\frac{\partial^2 I}{\partial r^2}+\hbar g(r)\left(\frac{f'(r)}{2f(r)}+\frac{g'(r)}{2g(r)}\right)\frac{\partial I}{\partial r}=0~.
\end{split}
\end{equation}
It can be inferred from the forms of eq.(s)(\ref{1.3},\ref{1.6}) that we can separate the equations involving the time coordinate ($t$) and the radial coordinate ($r$). Therefore, we can consider the form of $I(r,t)$ to be
\begin{equation}\label{1.7}
I(t,r)=\varepsilon t+\tilde{S}(r)
\end{equation}
where $\varepsilon$ denotes the energy of the particle. 

\noindent Now substituting the form of $I(r,t)$ from eq.(\ref{1.7}) in eq.(\ref{1.6}), we get
\begin{equation}\label{1.9}
\begin{split}
&\frac{i\varepsilon^2}{f(r)}-ig(r)\left(\frac{\partial\tilde{S}}{\partial r}\right)^2+\hbar g(r) \frac{\partial^2\tilde{S}}{\partial r^2}\\&+\hbar g(r)\left(\frac{f'(r)}{2f(r)}+\frac{g'(r)}{2g(r)}\right)\frac{\partial \tilde{S}}{\partial r}=0~.
\end{split}
\end{equation}
We now expand $\tilde{S}(r)$ in a power series in $\hbar$ as
\begin{equation}\label{1.8}
\tilde{S}(r)=S_0(r)+\hbar S_1(r)+\hbar^2S_2(r)+\cdots.
\end{equation}
Using this form of $\tilde{S}(r)$ from eq.(\ref{1.8}), we can recast eq.(\ref{1.9}) in the following form
\begin{equation}\label{1.11}
\begin{split}
&\frac{i\varepsilon^2}{f(r)}-ig(r)\biggr(\frac{\partial S_0}{\partial r}\biggr)^2+\hbar g(r)\biggr[-2i\frac{\partial S_0}{\partial r} \frac{\partial S_1}{\partial r}+\frac{\partial^2S_0}{\partial r^2}\\&+\left[\frac{f'(r)}{2f(r)}+\frac{g'(r)}{2g(r)}\right]\frac{\partial S_0}{\partial r}\biggr]+\hbar^2g(r)\biggr[-2i\frac{\partial S_0}{\partial r}\frac{\partial S_2}{\partial r}
\\&-i\left(\frac{\partial S_1}{\partial r}\right)^2+\frac{\partial^2 S_1}{\partial r^2}+\left[\frac{f'(r)}{2f(r)}+\frac{g'(r)}{2g(r)}\right]\frac{\partial S_1}{\partial r}\biggr]+\cdots=0~.
\end{split}
\end{equation}
Equating the terms involving equal powers of the reduced Planck's constant ($\hbar$) from eq.(\ref{1.11}) to zero, we obtain the following set of equations
\begin{align}
\hbar^0:&~ \frac{i\varepsilon^2}{f(r)}-ig(r)\biggr(\frac{\partial S_0}{\partial r}\biggr)^2=0~,\label{1.12}\\
\hbar^1:&~-2i\frac{\partial S_0}{\partial r} \frac{\partial S_1}{\partial r}+\frac{\partial^2S_0}{\partial r^2}+\left[\frac{f'(r)}{2f(r)}+\frac{g'(r)}{2g(r)}\right]\frac{\partial S_0}{\partial r}=0~,\label{1.13}\\
\hbar^2:&~-2i\frac{\partial S_0}{\partial r}\frac{\partial S_2}{\partial r}-i\left(\frac{\partial S_1}{\partial r}\right)^2+\left[\frac{f'(r)}{2f(r)}+\frac{g'(r)}{2g(r)}\right]\frac{\partial S_1}{\partial r}\biggr]\label{1.14}\\&+\frac{\partial^2 S_1}{\partial r^2}\nonumber=0~.\\
\vdots~~~&\nonumber
\end{align}
In order to obtain the complete form of $\tilde{S}(r)$ in eq.(\ref{1.8}), we need to solve the complete set of equations given by eq.(s)(\ref{1.12}-\ref{1.14}) and the higher order equations in $\hbar$ as well. From eq.(\ref{1.12}), we obtain the following solution
\begin{equation}\label{1.15}
\frac{\partial S_0}{\partial r}=\pm \frac{\varepsilon}{\sqrt{f(r)g(r)}}~.
\end{equation}
Integrating the above equation, with respect to the radial coordinate, we obtain
\begin{equation}\label{1.16}
S_0=\pm\varepsilon\int^r\frac{dr}{\sqrt{f(r)g(r)}}~.
\end{equation}
Substituting the form of $\frac{\partial S_0}{\partial r}$ from eq.(\ref{1.15}) in eq.(\ref{1.13}), we obtain the following relation
\begin{equation}\label{1.17}
\frac{\partial S_1}{\partial r}=0~.
\end{equation}
Eq.(\ref{1.17}) implies that $S_1$=constant. Substituting eq.(\ref{1.17}) back in eq.(\ref{1.14}), we obtain
\begin{equation}\label{1.18}
\frac{\partial S_2}{\partial r}=0~.
\end{equation}
Following a similar procedure, we find that
\begin{equation}\label{1.19}
\frac{\partial S_n}{\partial r}=0~,~\forall~n\in\{1,2,3,\ldots\}~.
\end{equation}
Eq.(\ref{1.19}) indicates that $S_n~$=~constant $\forall~n\in\{1,2,3,\ldots\}$ and therefore only $S_0$ survives while calculating the energy of the system. Using eq.(\ref{1.16}) in eq.(\ref{1.7}), we obtain
\begin{equation}\label{1.21}
I(t,r)=\varepsilon t\pm\int\frac{\varepsilon}{\sqrt{f(r)g(r)}}~.
\end{equation}
Using the form of $I(t,r)$ from eq.(\ref{1.21}), the solution for the massless scalar field takes the form
\begin{equation}\label{1.22}
\Psi=\exp\left(-\frac{i\varepsilon t}{\hbar}\mp \frac{i\varepsilon}{\hbar}\int\frac{dr}{\sqrt{f(r)g(r)}}\right)~.
\end{equation}
In the next section, we will be considering reflection of the massless scalar field from the event horizon of a quantum corrected black hole.
\section{Reflection from the horizon of a quantum corrected black hole}\label{Sec3}
The metric of a quantum corrected black hole in $(3+1)$-dimensions (obtained from the flow of the Newton's gravitational constant) reads\cite{Reuter0,Wetterich,ReuterWetterich,SReuter,Reuter}
\begin{equation}\label{1.23}
ds^2=-f(r)dt^2+\frac{dr^2}{f(r)}+r^2d\theta^2+r^2\sin^2\theta d\phi^2
\end{equation}
where
\begin{equation}\label{1.24}
f(r)=1-\frac{2G(r)M}{r}
\end{equation}
and the form of $G(r)$ is given by (in natural units)
\begin{equation}\label{1.25}
G(r)=\frac{G}{1+\frac{\tilde{\omega}G}{r^2}}~.
\end{equation}
In eq.(\ref{1.25}), $\tilde{\omega}$ is a constant denoting the quantum gravity corrections  to the black geometry . From the lapse function in eq.(\ref{1.24}), we obtain the inner and external radius of this quantum corrected black hole geometry as follows
\begin{equation}\label{1.26}
r_{\pm}=GM\pm\sqrt{G^2M^2-\tilde{\omega}G}~.
\end{equation}
In case of the of the black holes with $f(r)=g(r)$, the scalar field solution in eq.(\ref{1.22}) reduce to the following form
\begin{equation}\label{1.27}
\Psi=\exp\left(-\frac{i\varepsilon t}{\hbar}\mp \frac{i\varepsilon}{\hbar}\int\frac{dr}{f(r)}\right)~.
\end{equation}
In eq.(\ref{1.27}), $\int^r\frac{dr}{f(r)}=r_*$ denotes the tortoise coordinate. For the metric structure in eq.(\ref{1.24}), we obtain the form of the tortoise coordinate in the vicinity of the event horizon radius as
\begin{equation}\label{1.28}
\begin{split}
\int_{r_+}^r\frac{dr}{f(r)}&\cong GM\ln\left[r^2-2GMr+\tilde{\omega}G\right]+\\&\frac{G^2M^2}{\sqrt{G^2M^2-\tilde{\omega}G}}\ln\left[\frac{r-GM-\sqrt{G^2M^2-\tilde{\omega}G}}{r-GM+\sqrt{G^2M^2-\tilde{\omega}G}}\right]~.
\end{split}
\end{equation}
Using the forms of $r_\pm$ from eq.(\ref{1.26}), we can recast the above equation as
\begin{equation}\label{1.29}
\int_{r_+}^r\frac{dr}{f(r)}\cong \frac{2GM}{r_+-r_-}\left(r_+\ln\left[r-r_+\right]-r_-\ln\left[r-r_-\right]\right)~.
\end{equation}
The above form has two parts, one part indicates a solution inside the external radius $r_+$ and the other outside it. Here we will be considering the $\ln(r-r_+)$ part of the solution describing the scalar field solution outside the horizon radius $r_+$ as this is the only piece relevant for obtaining the Hawking temperature. Hence, the complete solution of the massless scalar field in the vicinity of the horizon radius $r_+$ is given by (ignoring the second term in eq.(\ref{1.29}))\footnote{\noindent A complete analysis keeping both the terms in eq.(\ref{1.29}) is given in an Appendix.}
\begin{equation}\label{1.30}
\Psi(t,r)=e^{-\frac{i\varepsilon t}{\hbar}}\phi(r)
\end{equation}
where $\phi(r)$ has the form given by
\begin{equation}\label{1.31}
\phi(r)=\exp\left[{\mp\frac{2i\varepsilon GM}{\hbar}\left(\frac{r_+}{r_+-r_-}\right)\ln(r-r_+)}\right]~.
\end{equation}
The `minus' sign in the wavefunction in eq.(\ref{1.31}) represents a radially infalling massless scalar field that completely gets absorbed by the event horizon of the black hole at $r=r_+$. In our case, we will be considering a radially outgoing solution as well in order to allow interference between the infalling and outgoing parts of the wavefunction. Therefore, the wavefunction in the vicinity of the horizon radius $r_+$ can be written as
\begin{equation}\label{1.32}
\Phi(r)=e^{-\frac{2i\varepsilon GM}{\hbar}\left(\frac{r_+\ln(r-r_+)}{r_+-r_-}\right)}+\mathcal{R}e^{\frac{2i\varepsilon GM}{\hbar}\left(\frac{r_+\ln(r-r_+)}{r_+-r_-}\right)}.
\end{equation}
The first term in eq.(\ref{1.32}) denotes an incoming wave whereas the outgoing wave is described by the second term. In case of interference between the incoming and the outgoing waves, the reflection coefficient $\mathcal{R}$ must have non-zero value. If the black horizon acts as a perfect absorber then the reflection coefficient $\mathcal{R}$ must be equal to zero. From the unitarity condition, we know that $|\mathcal{R}|\leq 1$. The wave function in eq.(\ref{1.32}) is singular at the point $r=r_+$. In order to tackle this issue, we shall take recourse to analytic continuation by considering $r-r_+=\mathcal{b}$ as a complex number. Then $\mathcal{b}$ can be expressed as
\begin{equation}\label{1.33}
\mathcal{b}=r-r_+=\mathcal{a}e^{i\varphi}
\end{equation} 
where $|\mathcal{b}|=\mathcal{a}$ is a real number and $\varphi$ denotes the argument of the complex number $\mathcal{b}$. Using eq.(\ref{1.33}), we can recast eq.(\ref{1.32}) as
\begin{equation}\label{1.34}
\Phi(r)=e^{-\frac{2i\varepsilon GM}{\hbar}\left(\frac{r_+}{r_+-r_-}\right)\ln\mathcal{b}}+\mathcal{R}e^{\frac{2i\varepsilon GM}{\hbar}\left(\frac{r_+}{r_+-r_-}\right)\ln\mathcal{b}}.
\end{equation} 
In order to proceed further we will now use the analytical continuation method \cite{Kuchiev}. In our analysis we are considering the massless scalar field very near the outer horizon $r_+$ of the quantum corrected black hole. Therefore, we can consider, $0<|\mathcal{b}|\ll1$. We will now continue to rotate the complex number $\mathcal{b}$ in the complex plane by an angle $2\pi$ (clockwise rotation). Under this clockwise rotation, the complex number $\mathcal{b}$ is given by the following relation
\begin{equation}\label{1.35}
\mathcal{b}_{2\pi}=\mathcal{\alpha}e^{i(\varphi-2\pi)}~.
\end{equation}
Using modified complex number from the above equation, the modified wave function $\Phi_{2\pi}(r)$ is given by
\begin{equation}\label{1.36}
\begin{split}
\Phi_{2\pi}(r)=&e^{-\frac{2i\varepsilon GM}{\hbar}\left(\frac{r_+}{r_+-r_-}\right)\ln\mathcal{b}_{2\pi}}+\mathcal{R}e^{\frac{2i\varepsilon GM}{\hbar}\left(\frac{r_+}{r_+-r_-}\right)\ln\mathcal{b}_{2\pi}}\\
=&\xi e^{-\frac{2i\varepsilon GM}{\hbar}\left(\frac{r_+}{r_+-r_-}\right)\ln\mathcal{b}}+\frac{\mathcal{R}}{\xi}e^{\frac{2i\varepsilon GM}{\hbar}\left(\frac{r_+}{r_+-r_-}\right)\ln\mathcal{b}}
\end{split}
\end{equation} 
where $\xi$ is given by
\begin{equation}\label{1.37}
\xi=\exp\left[-\frac{4\pi\varepsilon GM}{\hbar}\left(\frac{r_+}{r_+-r_-}\right)\right]~.
\end{equation}
The real differential equation satisfied by $\Phi(r)$ is also satisfied by the analytically continued function $\Phi_{2\pi}(r)$. This condition implies that one of the coefficients must have an absolute value equal to unity. In case of $\Phi(r)$ in eq.(\ref{1.34}), the first coefficient has an absolute value one. Now for the rotated wave function $\Phi_{2\pi}(r)$ the first coefficient $\xi=\exp\left[-\frac{4\pi\varepsilon GM}{\hbar}\left(\frac{r_+}{r_+-r_-}\right)\right]<1$, therefore the second coefficient must have an absolute value equal to one. Hence, we can write
\begin{equation}\label{1.38}
\begin{split}
\frac{|\mathcal{R}|}{\xi}&=1\\
\implies |\mathcal{R}|&=\xi=\exp\left[-\frac{4\pi\varepsilon GM}{\hbar}\left(\frac{r_+}{r_+-r_-}\right)\right]~.
\end{split}
\end{equation}
Therefore, we observe that the value of the reflection coefficient $\mathcal{R}$ is non-zero. Hence, the probability of reflection from the even horizon is given by
\begin{equation}\label{1.39}
\mathcal{P}=|\mathcal{R}|^2=\exp\left[-\frac{8\pi\varepsilon GM}{\hbar}\left(\frac{r_+}{r_+-r_-}\right)\right]~.
\end{equation} 
Now using the principle of detailed balance we can write
\begin{equation}\label{1.40}
\mathcal{P}=\exp\left[-\frac{\varepsilon}{k_B T}\right]=\exp\left[-\frac{8\pi\varepsilon GM}{\hbar c^3}\left(\frac{r_+}{r_+-r_-}\right)\right]
\end{equation}
where $k_B$ denotes the Boltzmann constant and $T$ denotes the Hawking temperature of the black hole. From eq.(\ref{1.40}), we can obtain the Hawking temperature for this quantum corrected black hole to be 
\begin{equation}\label{1.41}
\begin{split}
T&=\frac{\hbar c^3}{8\pi k_BGM}\left(1-\frac{r_-}{r_+}\right)\\
&=\frac{\hbar c^3}{8\pi k_BGM}-\frac{\hbar c^3}{8\pi k_BGM}\left[\frac{1-\sqrt{1-\frac{\hbar\tilde{\omega}c}{GM^2}}}{1+\sqrt{1-\frac{\hbar\tilde{\omega}c}{GM^2}}}\right]~.
\end{split}
\end{equation}
In general, the quantum gravity correction term $\tilde{\omega}$ is very small, hence we can simplify the above result as
\begin{equation}\label{1.42}
T\cong\frac{\hbar c^3}{8\pi k_BGM}\left(1-\frac{\hbar\tilde{\omega}c}{4GM^2}-\frac{\hbar^2\tilde{\omega}^2c^2}{8G^2M^4}+\mathcal{O}(\tilde{\omega}^3)\right)~.
\end{equation}
From the form of eq.(\ref{1.42}), we can observe that the Hawking temperature picks up quantum gravity corrections only due to the underlying quantum gravitational nature of the background spacetime geometry. Note that we have now considered clockwise rotation only. In case of counter-clockwise rotation, we would have arrived at the following relation
\begin{equation}\label{1.43}
|\mathcal{R}|=\exp\left[\frac{4\pi\varepsilon GM}{\hbar c^3}\left(\frac{r_+}{r_+-r_-}\right)\right]~.
\end{equation}
Now from eq.(\ref{1.43}), we observe that $|\mathcal{R}|>1$ which is in contradiction with the unitarity condition ($|\mathcal{R}|\leq 1$). Hence, the counter-clockwise rotation is forbidden in the current analysis. 

\noindent In the next section we will calculate the Hawking temperature for a \textit{Garfinkle-Horowitz-Strominger} (GHS) black hole for which $f(r)\neq g(r)$. 
\section{Reflection from the horizon of a \textit{Garfinkle -Horowitz-Strominger} black hole}
A generalized spacetime geometry arises in case of low energy string theories. Among several low energy string theoretical actions, we consider the following action
\begin{equation}\label{1.44}
\mathcal{S}_{\mathcal{A}}=\int d^4x\sqrt{-g}e^{-2\phi}(-R-4(\nabla\phi)^2+F^2)
\end{equation}
where $\phi$ is the dilaton field and $F_{\mu\nu}$ is a Maxwell field associated with a $U(1)$ subgroup of $E_8\times E_8$ or $\text{Spin}(32)/Z_2$. Among the family of solutions of the low energy action given in eq.(\ref{1.44}), its charged black hole solution in $(1+1)$-dimensions is given by\cite{GHS_Original,GHS1,GHS2,SG0} 
\begin{equation}\label{1.45}
ds^2=-f(r)dt^2+g(r)^{-1}dr^2
\end{equation}
where $f(r)$ and $g(r)$ are given by
\begin{align}
f(r)&=\left(1-\frac{2Me^{\phi_0}}{r}\right)\left(1-\frac{Q^2e^{3\phi_0}}{Mr}\right)^{-1}~,\label{1.46}\\
g(r)&=\left(1-\frac{2Me^{\phi_0}}{r}\right)\left(1-\frac{Q^2e^{3\phi_0}}{Mr}\right)\label{1.47}
\end{align}
where $\phi_0$ is the asymptotic constant value of the dilaton field. This is the ``\textit{Garfinkle-Horowitz-Strominger}" (GHS) black hole in $(1+1)$-dimensions \cite{GHS_Original}. The event horizon radius for a GHS black hole is given by
\begin{equation}\label{1.48}
r_+=2Me^{\phi_0}~.
\end{equation}
Using the metric structure from eq.(s)(\ref{1.46},\ref{1.47}), we can compute $S_0$ from eq.(\ref{1.16}) in the vicinity of the horizon radius $r_+$ as
\begin{equation}\label{1.49}
\begin{split}
S_0&=\pm\varepsilon \int^r_{r_+} \frac{dr}{\sqrt{f(r)g(r)}}=\pm \varepsilon\int^r_{r_+}\frac{dr}{1-\frac{2Me^{\phi_0}}{r}}\\
&\cong\pm 2M\varepsilon e^{\phi_0}\ln(r-2Me^{\phi_0})=\pm \varepsilon r_+ \ln(r-r_+)~.
\end{split}
\end{equation}
As before, we will consider $r-r_+=\mathcal{b}_1=\mathcal{a}_1 e^{i\varphi_1}$ with $\mathcal{a}_1=|\mathcal{b}_1|$ and $\varphi_1=\text{arg}(\mathcal{b}_1)$. Using this condition along with eq.(\ref{1.49}), we can write the radial part of the wave function given by (considering reflection from the event horizon)
\begin{equation}\label{1.50}
\Phi^1(r)=e^{-\frac{i\varepsilon}{\hbar}r_+\ln\mathcal{b}_1}+\mathcal{R}e^{\frac{i\varepsilon}{\hbar}r_+\ln\mathcal{b}_1}~.
\end{equation}
Following the same analytical continuation method as before we do a clockwise rotation by an angle $2\pi$ in the complex plane. Under this clockwise rotation the radial part of the wave function $\Phi^1(r)$ transforms as follows
\begin{equation}\label{1.51}
\begin{split}
\Phi^1_{2\pi}(r)&=e^{-\frac{i\varepsilon}{\hbar}r_+\ln\left[\mathcal{b}_1e^{-2\pi i}\right]}+\mathcal{R}e^{\frac{i\varepsilon}{\hbar}r_+\ln\left[\mathcal{b}_1e^{-2\pi i}\right]}\\
&={\xi_1} e^{-\frac{i\varepsilon}{\hbar}r_+\ln\mathcal{b}_1}+\frac{\mathcal{R}}{\xi_1}e^{\frac{i\varepsilon}{\hbar}r_+\ln\mathcal{b}_1}
\end{split}
\end{equation} 
where $\xi_1$ is given by
\begin{equation}\label{1.52}
\xi_1=\exp\left({-\frac{2\pi\varepsilon}{\hbar}r_+}\right)~.
\end{equation}
We observe that the coefficient of the first term in eq.(\ref{1.51}) has a value less than one, therefore by the arguments used in the earlier section we can conclude the following
\begin{equation}\label{1.53}
|\mathcal{R}|=\xi_1=\exp\left[{-\frac{2\pi\varepsilon}{\hbar}r_+}\right]~.
\end{equation} 
From eq.(\ref{1.53}), we can calculate the probability of reflection from the horizon of the GHS black hole as follows
\begin{equation}\label{1.54}
\mathcal{P}=|\mathcal{R}|^2=\exp\left[{-\frac{8\pi\varepsilon GMe^{\phi_0}}{\hbar c^3}}\right]~.
\end{equation}
In order to obtain the Hawking temperature $T$, we need to use the principle of detailed balance as before. The final form of the Hawking temperature for a \textit{Garfinkle-Horowitz-Strominger} black hole is given by
\begin{equation}\label{1.55}
T=\frac{\hbar c^3}{8\pi k_B G M}e^{-\phi_0}~.
\end{equation}
From eq.(\ref{1.55}), we observe that for a GHS black hole the Hawking temperature has the usual form of the temperature multiplied by an exponential term. Now for a very small value of the asymptotic constant value of the dilaton field we can use $e^{-\phi_0}\approx 1-\phi_0+\mathcal{O}(\phi_0^2)$~. Hence, the Hawking temperature takes the form as follows
\begin{equation}\label{1.56}
T\cong\frac{\hbar c^3}{8\pi k_B G M}-\frac{\hbar c^3\phi_0}{8\pi k_B G M}+\mathcal{O}(\phi_0^2)~.
\end{equation} 
The form of the Hawking temperature in eq.(\ref{1.56}) has the form similar to that of a base temperature term along with higher order correction terms in $\phi_0$. 

\noindent Before concluding our discussion, we would like to point out that the same result for the Hawking temperature can be obtained  for a quantum corrected black hole geometry using the tunneling method \cite{PWZ} using the form of the action derived earlier in eq.(\ref{1.21}). The solution of the covariant Klein-Gordon equation from eq.(\ref{1.22}) can be decomposed into one ingoing solution and one outgoing solution as follows 
\begin{align}
\Psi_{in}=&\exp\left[-\frac{i\varepsilon t}{\hbar}-\frac{i\varepsilon}{\hbar}\int\frac{dr}{\sqrt{f(r)g(r)}}\right]\label{1.57a}\\
\Psi_{out}=&\exp\left[-\frac{i\varepsilon t}{\hbar}+\frac{i\varepsilon}{\hbar}\int\frac{dr}{\sqrt{f(r)g(r)}}\right]\label{1.58a}~.
\end{align}
Here we are considering the case of tunneling across the event horizon of the black hole. When the particle crosses the horizon, we observe an interchange of signs between the space and time coordinates indicating an imaginary part of the time coordinate for crossing the horizon. Two patches across the horizon are generally connected by an imaginary time\cite{Akhmedov}. Hence, by using the ingoing and outgoing wave functions in eq.(s)(\ref{1.57a},\ref{1.58a}), we can express the ingoing and outgoing solutions of the particle as \cite{PWZ}
\begin{align}
\mathcal{P}_{in}=&|\Psi_{in}|^2=\exp\left(\frac{2\varepsilon}{\hbar}\Im\left[t\right]+\frac{2\varepsilon}{\hbar}\Im\left[\int_0^r\frac{dr}{\sqrt{f(r)g(r)}}\right]\right)~,\label{1.59a}\\
\mathcal{P}_{out}=&|\Psi_{out}|^2=\exp\left(\frac{2\varepsilon}{\hbar}\Im\left[t\right]-\frac{2\varepsilon}{\hbar}\Im\left[\int_0^r\frac{dr}{\sqrt{f(r)g(r)}}\right]\right)\label{1.60a}~
\end{align}
where $\Im[t]$ denotes the imaginary part of time. Using the principle of detailed balance \cite{PWZ}, we can write the following
\begin{equation}\label{1.61a}
\begin{split}
\frac{\mathcal{P}_{out}}{\mathcal{P}_{in}}=\exp\left[-\frac{4\varepsilon}{\hbar}\Im\left[\int_0^r\frac{dr}{\sqrt{f(r)g(r)}}\right]\right]=\exp\left[-\frac{\varepsilon}{T}\right]~.
\end{split}
\end{equation}  
From eq.(\ref{1.61a}), we can extract the form of the Hawking temperature to be
\begin{equation}\label{1.62a}
T=\frac{\hbar}{4}\left(\Im\left[\int_0^r\frac{dr}{\sqrt{f(r)g(r)}}\right]\right)^{-1}~.
\end{equation}
The above equation gives the Hawking temperature for a large class of spherically symmetric black holes from the viewpoint of tunneling across the horizon. Substituting the form of the quantum corrected black hole geometry in eq.(\ref{1.62a}), we recover eq.(\ref{1.41}) thereby agreeing with the interference and reflection method. It is very important to notice that if the lapse function does not possess an explicit quantum gravity correction then the Hawking temperature also has no such corrections.
\section{Conclusion}
In this paper, we have used the method of \textit{reflection from the horizon} to investigate the Hawking temperature of a quantum corrected black hole. At first we have calculated the Hawking temperature of the quantum corrected black hole using the consideration that there will be a finite probability of a part of the scalar wave function being reflected from the event horizon of the black hole. The computed temperature formula for the quantum corrected black hole has structure similar to that of the Hawking temperature of a classical black hole (Schwarzschild black hole) along with higher order quantum gravity corrections. We have also computed the Hawking temperature for a GHS black hole. For the next part of our analysis, we assume tunneling through the event horizon of a black hole. We obtained the same result for the Hawking temperature of the quantum corrected black hole as obtained earlier using the method of \textit{reflection from the horizon}. These findings showcase the fact that quantum corrections arise in the Hawking temperature of the black hole only due to underlying quantum gravitational effects in the spacetime geometry. These results also indicate clearly that there should be no quantum gravity corrections in the Hawking temperature of the black hole if there are no such quantum gravitational effects in the geometry of the spacetime. 
\section*{Appendix: Hawking temperature with the complete solution in eq.(\ref{1.29})}
In section  (\ref{Sec3}), we have calculated the form of the Hawking temperature using the part of the solution outside the event horizon of the black hole in eq.(\ref{1.29}). In this Appendix, we shall proceed with the same calculation keeping both the terms. 
Following the discussion after eq.(\ref{1.30}), we can write the form of the radial part of the wavefunction in the vicinity of the event horizon keeping both the terms in eq.(\ref{1.26}) as follows
\begin{equation}\label{A2}
\begin{split}
\Phi(r)&=e^{-\frac{2i\varepsilon G M}{\hbar}\left(\frac{r_+}{(r_+-r_-)}\ln(r-r_+)-\frac{r_-}{(r_+-r_-)}\ln(r-r_-)\right)}\\&+\mathcal{R}e^{\frac{2i\varepsilon G M}{\hbar}\left(\frac{r_+}{(r_+-r_-)}\ln(r-r_+)-\frac{r_-}{(r_+-r_-)}\ln(r-r_-)\right)}~.
\end{split}
\end{equation}
Again defining $r-r_+=\mathcal{b}=\mathcal{a}e^{i\varphi}$, we can recast the above relation in the following form
\begin{equation}\label{A3}
\begin{split}
\Phi(r)&=e^{-\frac{2i\varepsilon G M}{\hbar}\left(\frac{r_+}{(r_+-r_-)}\ln\mathcal{b}-\frac{r_-}{(r_+-r_-)}\ln(\mathcal{b}+(r_+-r_-))\right)}\\&+\mathcal{R}e^{\frac{2i\varepsilon G M}{\hbar}\left(\frac{r_+}{(r_+-r_-)}\ln\mathcal{b}-\frac{r_-}{(r_+-r_-)}\ln(\mathcal{b}+(r_+-r_-))\right)}~.
\end{split}
\end{equation}
Using the same analytic continuation procedure \cite{Kuchiev} and rotating the complex number $\mathcal{b}$ by an angle $2\pi$ in the clockwise direction on the complex plane,  we can recast the rotated form of the radial part of the wavefunction as
\begin{equation}\label{A4}
\begin{split}
\Phi(r)&=e^{-\frac{2i\varepsilon G M}{\hbar}\left(\frac{r_+}{(r_+-r_-)}\ln\mathcal{b}_{2\pi}-\frac{r_-}{(r_+-r_-)}\ln(\mathcal{b}_{2\pi}+(r_+-r_-))\right)}\\&+\mathcal{R}e^{\frac{2i\varepsilon G M}{\hbar}\left(\frac{r_+}{(r_+-r_-)}\ln\mathcal{b}_{2\pi}-\frac{r_-}{(r_+-r_-)}\ln(\mathcal{b}_{2\pi}+(r_+-r_-))\right)}~.
\end{split}
\end{equation}
where the form of $\mathcal{b}_{2\pi}$ is given by eq.(\ref{1.35}). Since $\ln(\mathcal{b}_{2\pi}+(r_+-r_-))=\ln(\mathcal{b}e^{-2\pi i}+(r_+-r_-))=\ln(\mathcal{b}+(r_+-r_-))$, hence eq.(\ref{A4}) takes the form
\begin{equation}\label{A5}
\begin{split}
\Phi(r)&=\xi e^{-\frac{2i\varepsilon G M}{\hbar}\left(\frac{r_+}{(r_+-r_-)}\ln\mathcal{b}-\frac{r_-}{(r_+-r_-)}\ln(\mathcal{b}+(r_+-r_-))\right)}\\&+\frac{\mathcal{R}}{\xi}e^{\frac{2i\varepsilon G M}{\hbar}\left(\frac{r_+}{(r_+-r_-)}\ln\mathcal{b}_{2\pi}-\frac{r_-}{(r_+-r_-)}\ln(\mathcal{b}_{2\pi}+(r_+-r_-))\right)}
\end{split}
\end{equation}
where the analytical form of $\xi$ is given in eq.(\ref{1.37}).
By continuing the analysis in section (\ref{Sec3}) after eq.(\ref{1.37}), the same form of the Hawking temperature $T$ given in eq.(\ref{1.41}) is obtained. 

\end{document}